\documentclass{ws-procs9x6-cpt16}
\begin{document}

\newcommand{\refeq}[1]{(\ref{#1})}
\def\etal {{\it et al.}}

\title{Modeling the Dispersion and Polarization Content of\\ 
Gravitational Waves for Tests of General Relativity}

\author{Rhondale Tso,$^{1,2}$ Maximiliano Isi,$^{2}$ Yanbei Chen$^{1,2}$, and Leo Stein$^1$}

\address{$^1$Theoretical Astrophysics, California Institute of Technology\\
Pasadena, CA 91125, USA}

\address{$^2$LIGO Laboratory, California Institute of Technology,
Pasadena, CA 91125, USA}

\begin{abstract}
We propose a generic, phenomenological approach to modifying the dispersion of gravitational waves, independent of corrections to the generation mechanism.  This model-independent approach encapsulates all previously proposed parametrizations, including Lorentz violation in the Standard-Model Extension, and provides a roadmap for additional theories.  Furthermore, we present a general approach to include modulations to the gravitational-wave polarization content.  The framework developed here can be implemented in existing data analysis pipelines for future gravitational-wave observation runs.
\end{abstract}

\bodymatter

\section{Introduction}

Amendments to General Relativity (GR) in gravitational-wave (GW) physics have largely been applied to the generation mechanism, concentrating on the conservative and dissipative processes of the source.\cite{ppE}  
Modifications to GW dynamics can be applied by means of a small parameter $\epsilon_1$ introducing deviations to the GR emission process at the source, and a second, independent small parameter $\epsilon_2$, encoding modifications to the GW dispersion and propagation.  Here we consider the case where the $\epsilon_2$ correction, separate from $\epsilon_1$  controlling non-GR dynamics of the source,  primarily governs the correction.  Along its trajectory, GW fluctuations go through many cycles and accumulated effects could dominate over $\epsilon_1$, i.e., $\epsilon_2 > \epsilon_1$.  

Previous works have looked at massive graviton theories, modeled with a dispersion according to which GWs follow timelike trajectories.\cite{massive_graviton}   Later work then extended this method to include generic Lorentz violations mapped into the parametrized post-Einsteinian (ppE) framework.\cite{massive_graviton_LV}   Model-independent, Lorentz-violating frameworks have also been developed within the Standard-Model Extension (SME), to provide a description of frequency and anisotropic dependence.\cite{LV_SME}  

Using data from the first of GW detections, the Laser Interferometer Gravitational-Wave Observatory (LIGO) has recently imposed the most stringent bounds on the graviton wavelength, as well as other tests.\cite{tGR}  The goal of this project is to provide a framework to be directly implemented in LIGO parameter estimation routines.  

\subsection{Modified dispersion}

Here a generic alteration to the GW dispersion is performed, inspired by the electromagnetic analogue of dispersion in materials: $A_{\alpha \beta}(k^\mu) = 0$, with $k^\mu = (\omega, \vec{k})$ the GW four-momentum.  Expanding the general tensor $A_{\alpha \beta}$ on a flat background gives:
\begin{eqnarray}
A_{\alpha \beta }  &=&  A^{(0)}_{\alpha \beta } + i k^\lambda A^{(1)}_{\lambda \alpha \beta } + \frac{1}{2} k^\lambda k^\mu  A^{(2)}_{ \lambda \mu \alpha \beta} + i \frac{1}{3} k^\lambda k^\mu k^\nu  A^{(3)}_{\lambda \mu \nu \alpha \beta }+ \cdots . 
\end{eqnarray}   
Assuming that $A^{(n)}_{(\cdots) \alpha \beta } \propto M_{\alpha \beta }$, with $M_{\alpha \beta }$ an arbitrary constant, non-degenerate matrix, the above can  be reformulated as
\begin{equation}\label{nGR_dispersion}
-\omega^2 + |\vec{k}|^2 = G_0 + \hat{n}_j G^j_1 +  \hat{n}_i  \hat{n}_j G^{ij}_2 +  \hat{n}_i  \hat{n}_j \hat{n}_k G^{ijk}_3 +  \cdots ,
\end{equation}
where $\hat{n}$ is the direction of propagation and
\begin{eqnarray}
G_0 (\omega) &=& a + \omega b  + \omega^2 c + \cdots , \nonumber \\
G^j_1 (\omega)  &=& \left( a^j + \omega b^j + \omega^2 c^j + \cdots \right)  |\vec{k}| , \nonumber \\
G^{ij}_2 (\omega) &=& \left( a^{ij} + \omega b^{ij} + \omega^2 c^{ij} + \cdots \right) |\vec{k}|^2  ,\nonumber \\
G^{ijk}_3 (\omega) &=& \left( a^{ijk} + \omega b^{ijk} + \omega^2 c^{ijk} + \cdots \right) |\vec{k}|^3 \nonumber \\
\vdots && \vdots 
\end{eqnarray}
with coefficients $a, b, c , \dots \in \mathbb{C}$, etc.  Imaginary terms induce dissipation in the GW.  Study of the nonminimal, gravitational sector of the SME in Ref.\ \refcite{LV_SME} reveals operator $\hat{\bar{s}}_{\mu \nu}$ returning even powers of $k^\mu$ with no dissipation allowed.  Ignoring frequency dependence the isotropic limits of $a^{ij}, a^{ijk}, \dots$ with $a = - m_g^2$ returns models encompassed by the ppE framework.\cite{massive_graviton, massive_graviton_LV}

\subsection{Modified polarization}

Consideration of possible coordinate dependence of non-GR coefficients in Eq.\ \refeq{nGR_dispersion} motivates us to generically investigate polarization.  Such effects include linear and circular polarization.  Previous studies have considered such behaviors, effects including amplitude birefringence, extra degrees of freedom from bimetric theories, and birefringence from Lorentz violation in the SME.\cite{polarization}  Further work from the authors has also considered additional dynamics from non-GR polarizations with degrees of freedom propagating at speeds $v \neq 1$ and interacting with the GW's polarization content.\cite{future_paper}  

Specifically, from the SME, modifications to the polarization content is related to circular polarization.  Here the $+, \times$ modes can be written in terms of left- and right-handedness, with the emission $h_{L, R}$ rotated throughout its propagation and arriving as $h_{L, R}'$, 
\begin{equation}
h_{L, R} = h_+ \pm i h_\times, \phantom{aa} h_{L, R}' = \left( h_+ \pm i h_\times \right) e^{i k_{L, R} (\omega) \cdot D},
\end{equation}
where $k_{L, R} (\omega)$ stems from the dispersion Eq.\ \refeq{nGR_dispersion} and $D$ is the distance from observer to the source.  Accurate measurements of orbital inclination and distance from binary compact objects can provide accurate information of $+, \times$ polarization contents and the degree to which they are rotated.  

\section{Modified waveform}

For nondissipative coefficients in Eq.\ \refeq{nGR_dispersion} the modified waveform can be computed by considering the group velocity of GWs and looking at the difference in arrival time between wave packets emitted with delay $\Delta t_e$,
\begin{equation}
\Delta t_a = \Delta t_e ( 1 + z ) + \int \frac{dt}{a(t)} \left( \delta_\omega ( t; \omega_a ) - \delta_\omega ( t; \omega_a' )  \right).
\end{equation}
Here $\Delta t_a$ is the delay in arrival of two wave packets, while the dimensionless parameter $\delta_\omega$ encodes modifications to the dispersion assuming small departures from GR.  Also, $a(t)$ is the cosmological expansion parameter, $z$ the redshift, $\omega_a$ is the GW frequency at arrival  with primed quantities corresponding to the second emitted wave packet.  Note that $\delta_\omega$ comes from the implicit solution of the polynomial of Eq.\ \refeq{nGR_dispersion} for $\omega$.

\begin{figure}[t]
\includegraphics[width=2.2in]{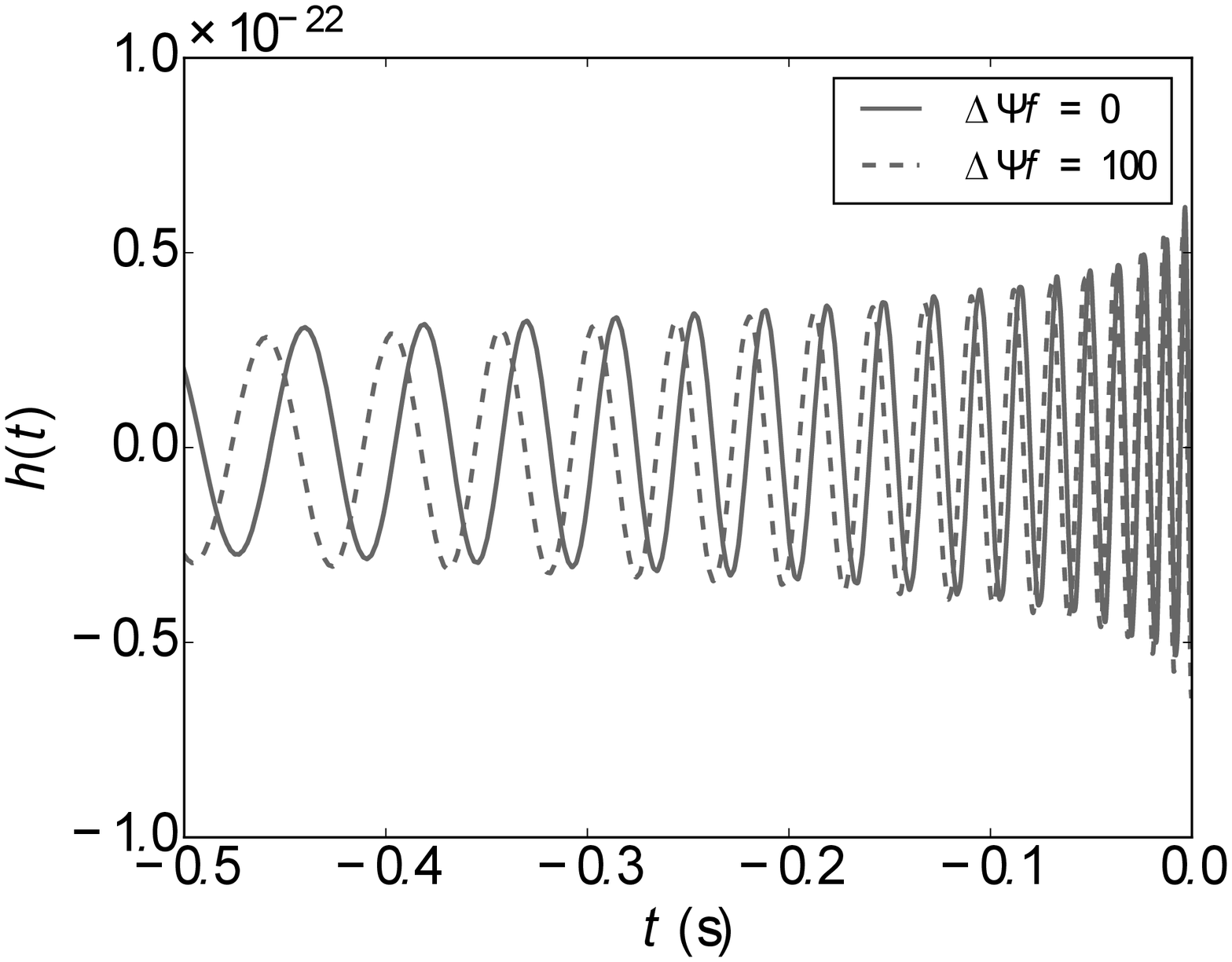}
\includegraphics[width=2.12in]{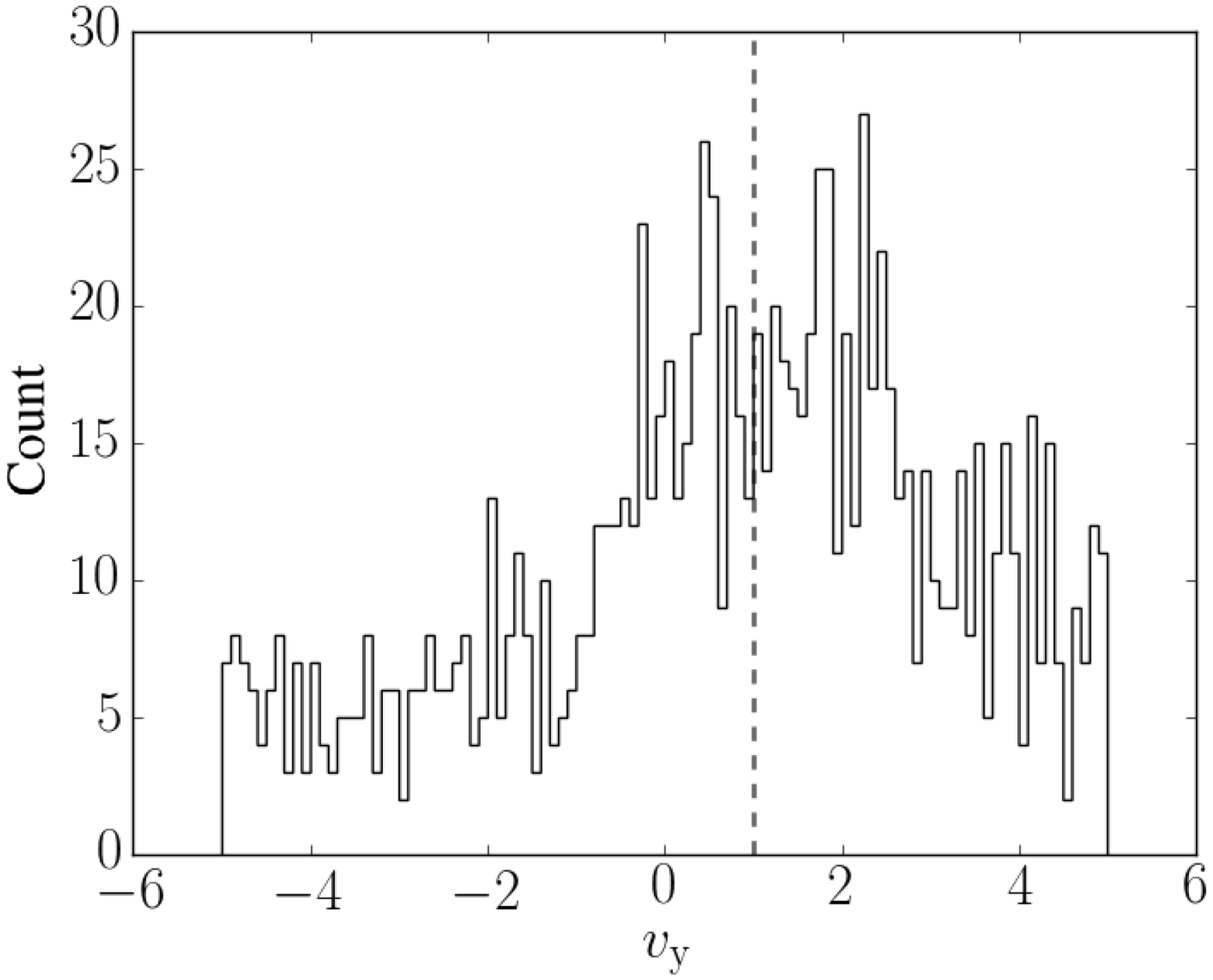}
\caption{Left: IMR signal of mock event for our toy model.  The solid line represents the GR limit, while the dashed line corresponds to non-GR modifications.  Right: Unnormalized posteriors for $v_y$ projection for event generated from mock data with IMR PhenomPv2 of no spin assuming Advanced LIGO noise.  The results are generated when the source location is known exactly; the distance is set to $410$ Mpc. }
\label{mod_waveform}
\end{figure}

This frequency dependent delay $\Delta t_a$ can be translated into a phase shift.  For a waveform $\tilde{h}(f) = A(f) \exp[i \Psi(f)]$, the correction for nondissipative terms will be $\Psi(f) \rightarrow  \Psi_{\mathrm{GR}}(f) + \Delta \Psi(f)$, where
\begin{equation}
\Delta \Psi(f) = \int\limits_{f_c}^{f} \int\limits_{t_e}^{t_a} dt d\tilde{f}  \frac{2 \pi}{a(t)} \left( \delta_f(t; \tilde{f}) - \delta_f(t; f_c)  \right)
\end{equation}
encapsulates the non-GR effects arising from the modified dispersion, where we have made the substitution $f = \omega / 2 \pi$ and $f_c$ is the coalescing frequency when considering compact binaries.  As a demonstration the left panel of Fig.\ \ref{mod_waveform} displays an inspiral-merger-ringdown (IMR) waveform with the extra phase shift arising from a modified dispersion of the form $-\omega^2 + | \vec{k}|^2 = - (m_g^2 + \hat{n} \cdot \vec{v}) $, with $\hat{n}$ the wave's direction of propagation and $\vec{v}$ an arbitrary vector.  The non-GR effects are largely exaggerated.  The massive graviton and anisotropic terms are degenerate since they both present dependence $\Delta \Psi \propto D/f$.  This exemplifies degeneracies that may exist in our dispersion Eq.\  \refeq{nGR_dispersion} and can be broken by coherently analyzing multiple detections.  The right panel of Fig.\ \ref{mod_waveform} displays an example of an unnormalized posterior distribution of $v_y$, the projection of the anisotropic GR-violating term appearing in the modified dispersion with the dashed line marking the injected value.  Here, $\hat{x}\equiv$ vernal equinox, $\hat{z}\equiv$ celestial north pole, and $\hat{y} = \hat{z} \times \hat{x}$.  How well each component $(v_x, v_y, v_z)$ can be measured depends on the location of the source. 

\end{document}